%% file: main.tex
\documentclass[10pt,conference]{IEEEtran}\IEEEoverridecommandlockouts
\usepackage{cite}
\usepackage{amsmath,amssymb,amsfonts}
\usepackage{algorithm, algpseudocode}
\algnewcommand{\LineComment}[1]{\Statex \hspace{\algorithmicindent}\(\triangleright\) \textit{#1}}

\usepackage{tcolorbox}
\usepackage{xcolor}
\usepackage{booktabs}
\usepackage{array}
\tcbuselibrary{skins, breakable}
 
\definecolor{primaryblue}{HTML}{D6E8F7}
\definecolor{secondaryred}{HTML}{FAD9D5}
\definecolor{responsegreen}{HTML}{DFF2E1}
\definecolor{responseorange}{HTML}{FDEBD0}
\definecolor{labelblue}{HTML}{1A5276}
\definecolor{labelred}{HTML}{922B21}
\definecolor{labelgreen}{HTML}{1E8449}
\definecolor{labelorange}{HTML}{B7770D}
\definecolor{boxborder}{HTML}{AAAAAA}
 
\tcbset{
  conversationbox/.style={
    enhanced,
    breakable,
    colframe=boxborder,
    colback=white,
    boxrule=0.4pt,
    arc=3pt,
    left=6pt, right=6pt, top=4pt, bottom=4pt,
    fonttitle=\small\bfseries,
    attach boxed title to top left={yshift=-2mm, xshift=4mm},
    boxed title style={colback=white, colframe=boxborder, boxrule=0.4pt, arc=2pt}
  },
  primarybox/.style={
    colback=primaryblue, colframe=labelblue,
    boxrule=0.5pt, arc=2pt,
    left=4pt, right=4pt, top=2pt, bottom=2pt,
    fontupper=\small
  },
  secondarybox/.style={
    colback=secondaryred, colframe=labelred,
    boxrule=0.5pt, arc=2pt,
    left=4pt, right=4pt, top=2pt, bottom=2pt,
    fontupper=\small
  },
  respnosuffixbox/.style={
    colback=responsegreen, colframe=labelgreen,
    boxrule=0.5pt, arc=2pt,
    left=4pt, right=4pt, top=2pt, bottom=2pt,
    fontupper=\small
  },
  respwithsuffixbox/.style={
    colback=responseorange, colframe=labelorange,
    boxrule=0.5pt, arc=2pt,
    left=4pt, right=4pt, top=2pt, bottom=2pt,
    fontupper=\small
  },
}
 
\newcommand{\rolelabel}[2]{%
  \noindent{\footnotesize\textcolor{#1}{\textbf{#2}}}\\[1pt]%
}

 \usepackage{amsmath}
\usepackage{amssymb}
\usepackage{graphicx}
\usepackage{textcomp}
\usepackage{xcolor}
\usepackage{hyperref}
\usepackage{multirow}
\usepackage{float}
\usepackage{makecell}
\usepackage{tikz}
\usetikzlibrary{arrows.meta,positioning,fit,calc}

\def\BibTeX{{\rm B\kern-.05em{\sc i\kern-.025em b}\kern-.08em
    T\kern-.1667em\lower.7ex\hbox{E}\kern-.125emX}}
\begin{document}

\title{Bypassing Prompt Injection Detectors through Evasive Injections}

\author{\IEEEauthorblockN{ JR Jahed,  Ihsen Alouani}
\IEEEauthorblockA{\textit{Centre for Secure Information Technologies} \\
\textit{EEECS, Queen's University Belfast}\\
Belfast, UK \\
mrahman13@qub.ac.uk \hspace{10pt} i.alouani@qub.ac.uk}}

\maketitle

\begin{abstract}
Large language models (LLMs) are increasingly used in interactive and retrieval-augmented systems, but they remain vulnerable to prompt injection attacks, where injected secondary prompts force the model to deviate from the user’s instructions to execute a potentially malicious task defined by the adversary. Recent work shows that ML models trained on activation shifts from LLMs' hidden layers can detect such drift. In this paper, we demonstrate that these detectors are not robust to adaptive
adversaries. We propose a multi-probe evasion attack that appends an adversarially optimised
suffix to poisoned inputs, jointly optimising a universal suffix to
simultaneously fool all layer-wise drift detectors while preserving the
effectiveness of the underlying injection.
 Using a modified Greedy Coordinate Gradient (GCG) approach, we generate universal suffixes that make prompt injections consistently evasive across multiple probes simultaneously. On Phi-3 3.8B and Llama-3 8B, a single suffix achieves attack success rates of 93.91\% and 99.63\% in successfully evading all detectors simultaneously. These results show that activation-based task drift detectors are highly vulnerable to adaptive prompt injection attacks, motivating stronger defences against such threats. We also propose a defence based on adversarial suffix augmentation: we generate multiple suffixes, append one at random during forward passes, and train detectors on the resulting activations. This approach is found to be effective against evasive attacks. \thanks{Code available at \href{https://github.com/ihsenLab/Evasive-Prompt-Injection-IJCNN2026-}{\texttt{[Github]}}}
\end{abstract}


\section{Introduction}

Large language models (LLMs) are effectively fuelling a paradigm shift in digital automation owing to their capacity to follow natural language instructions and to execute a wide variety of tasks. However, with this increasingly widespread deployment, they also became commonplace for a myriad of attacks \cite{zaree-etal-2025-attention,glvlsi,greshake2023not}. Particularly, LLMs are increasingly deployed in retrieval-augmented generation (RAG) systems~\cite{lewis2020retrieval,jiang2023active}, where they answer user queries by incorporating external data retrieved from databases, documents, or the web. This architecture introduces a new threat surface: the retrieved data, which originates from potentially untrusted sources, may contain adversarially crafted instructions. When an LLM processes such \emph{poisoned} data, it may execute the injected instruction instead of, or in addition to, the user's intended query, a phenomenon known as \emph{prompt injection}~\cite{greshake2023not,perez2022ignore}.

Prompt injection poses a serious threat to the integrity and reliability of LLM-based systems. An attacker who controls even a small portion of the data that an LLM retrieves can hijack the model's behaviour: exfiltrating sensitive information, generating misleading outputs, or executing arbitrary instructions. As LLMs become embedded in high-stakes applications, customer service, code generation, enterprise search, the consequences of successful prompt injection grow increasingly severe. The core challenge is that LLMs struggle to distinguish between legitimate user instructions and malicious instructions embedded in data, violating the principle of data-instruction separation.

Several defences have been proposed to mitigate prompt injection. Input filtering and preprocessing attempt to detect and remove suspicious patterns before they reach the model~\cite{liu2024formalizing}. Instruction hierarchy approaches train models to prioritise certain instruction sources over others. Perplexity-based detection flags inputs that deviate from expected distributions. However, these defences often fail against sophisticated attacks or impose significant usability constraints.

A promising recent approach by Abdelnabi et al.~\cite{abdelnabi2025get} detects prompt injection by monitoring the LLM's internal representations. Their key insight is that when an LLM begins executing an injected instruction, deviating from the user's original task, its internal activations shift in detectable ways. They train lightweight logistic regression classifiers (linear probes) on the \emph{activation deltas} at multiple hidden layers, capturing the difference between how the model processes clean versus poisoned inputs. This approach achieves high detection accuracy against standard prompt injection attacks and offers an attractive property: it operates on fixed internal representations without modifying the LLM itself.

However, the robustness of activation-based detection under \emph{adaptive} adversaries remains unexplored. A natural question arises: \textbf{can an attacker craft inputs that successfully inject a malicious prompt while simultaneously evading the activation-based detectors?}

In this paper, we demonstrate that such evasion is \textit{not only possible but highly effective}. Our key insight is that the attacker's optimisation objective can be \emph{decoupled}: the poisoned prompt carries the malicious instruction that hijacks the LLM's behaviour, while an \emph{adversarial suffix}, a carefully optimised sequence of tokens appended to the input, manipulates the model's activations to evade detection. Critically, this suffix does not need to alter the semantic content of the injection; it only needs to shift the activation patterns at the monitored layers into regions that the linear probes classify as benign.

We formulate this as the following problem: We try to find a suffix that simultaneously fools all layer-wise probes while preserving the effectiveness of the underlying prompt injection. To solve this problem, we adapt and repurpose the Greedy Coordinate Gradient (GCG) algorithm~\cite{zou2023universal}, originally developed as a jailbreak attack on aligned LLMs, to target the detection mechanism instead. By back-propagating through the LLM and injecting gradients from each probe's loss function, we optimise suffix tokens that induce activations misclassified as clean by all detectors simultaneously.

Our experiments on Phi-3 3.8B and Llama-3 8B reveal that activation-delta probes are highly vulnerable to this attack. A single universal suffix, one that transfers across thousands of different poisoned prompts, achieves attack success rates of 93.91\% and 99.63\% respectively when all probes must be fooled, and near-perfect evasion ($>$99\%) under majority voting. These results expose a significant gap between detection performance on standard benchmarks and robustness against adaptive adversaries.

To address this vulnerability, we propose a defence based on \emph{adversarial suffix augmentation}: we generate a diverse pool of adversarial suffixes, randomly append them to training data, and retrain the probes on the resulting activations. This exposes the detectors to the distribution of adversarial activation patterns, making them robust to novel suffixes at test time. Our suffix-augmented probes achieve 80--100\% detection accuracy against held-out adversarial suffixes while maintaining performance on clean inputs.

\noindent\textbf{Contributions.} We make the following contributions:
\begin{itemize}
    \item We develop a multi-probe evasion attack that jointly optimises an adversarial suffix to fool all layer-wise task drift detectors simultaneously, achieving $>$93\% evasion on Phi-3 and $>$99\% on Llama-3.
    \item We propose randomised suffix augmentation as a defence, which proves highly resistant to both known and novel adversarial suffixes.
    \item We provide the first systematic evaluation of activation-based prompt injection detectors under adaptive threat models, revealing fundamental limitations of activation space ML detectors.
\end{itemize}
\section{Background}

\textbf{Prompt Injection Attacks:} Prompt injection is a general class of attacks against large language models (LLMs), where an attacker appends or embeds malicious instructions into model inputs \cite{perez2022ignore}, \cite{greshake2023not}. Instead of following the user’s intended task, the LLM may then execute the injected instruction. This undermines the integrity of the system and can lead to unintended or harmful behaviours. Prompt injection can occur in many contexts, ranging from direct user inputs to data retrieved from external sources. Retrieval-augmented generation (RAG) systems are especially vulnerable, since retrieved data can be attacker-controlled, allowing secondary instructions to be stealthily injected into the context. Such injections violate the principle of data–instruction separation, as contextual data should be used only to support the user’s task and never be executed as instructions.

\textbf{Adversarial Attacks on LLMs: } Adversarial examples were first studied in computer vision, where imperceptible perturbations to input images could cause classifiers to mispredict \cite{goodfellow2014explaining, Tarchoun_2023_CVPR}. Extending these ideas to natural language processing (NLP) is more challenging due to the discrete nature of tokens and the requirement to preserve semantic meaning. Nonetheless, a wide range of adversarial strategies have been proposed, including synonym substitution, paraphrasing, character-level perturbations, and suffix-based attacks. In the context of LLMs, adversarial suffixes are especially investigated in jailbreak attacks: an attacker appends carefully optimised sequences of tokens to a prompt to manipulate an aligned model’s behaviour and force it to generate harmful content. One method for generating such suffixes is the Greedy Coordinate Gradient (GCG) method \cite{zou2023universal}, which iteratively updates suffix tokens by estimating gradients and selecting the most promising replacements.

\textbf{Task Drift Detection with Linear Probes:} To defend against prompt injection, Abdelnabi et al. \cite{abdelnabi2025get} introduced the concept of task drift detection. Their method attaches logistic regression classifiers (linear probes) to multiple hidden layers of an LLM, trained on the activation deltas between clean and poisoned inputs. The intuition is that if a model begins to follow a secondary instruction, the internal activations diverge from those induced by the user’s primary instruction. By monitoring these activation deltas, the probes can distinguish clean prompts from poisoned ones with high accuracy. This approach provides a lightweight and model-agnostic detection mechanism, as the probes are simple linear classifiers operating on fixed representations.

Our work builds directly on this framework. While Abdelnabi et al. \cite{abdelnabi2025get} evaluated robustness against straightforward prompt injection, we investigate whether the same linear probes can be circumvented through adversarial optimisation of suffixes.

\section{Threat Model}
\subsection{System Model}



We assume a general-purpose retrieval-augmented language model (LLM), similar to the one described in \cite{abdelnabi2025get}. The user can instruct the LLM to perform a task, i.e., the primary instruction. The LLM requires and obtains external data to successfully execute user's instruction. This data can be classified into two categories: clean and poisoned. Clean data contains only factual or contextual information relevant to the user’s task. Poisoned data, in contrast, contains an injected secondary instruction intended to influence the LLM. This injected instruction can be either malicious (e.g., “how to make a bomb?”) or benign (e.g., “what is the White House?”). Even a benign instruction is unwanted in this context because of the violation of data-instructions separation.


Because the retrieved data originates from external, potentially untrusted sources such as email in the victim mailbox, we assume the attacker controls the data channel. The attacker can therefore inject secondary instructions into the retrieved text. However, the attacker cannot modify the user’s primary instruction, nor can they alter the parameters of the LLM or the task drift detectors.

We distinguish two attacker settings:

\begin{enumerate}
    \item \textbf{Baseline attacker (as in Abdelnabi et al. \cite{abdelnabi2025get}):}
    This attacker injects a secondary instruction into the retrieved text, with the goal of making the LLM execute it. This is the threat model considered in the original task drift detection work.

    \item \textbf{Adaptive attacker (our extended formulation):}
    We consider a stronger attacker who, in addition to injecting a secondary instruction, appends an adversarial suffix to the poisoned text. This suffix is optimised to fool the linear classifiers trained to detect task drift. By bypassing detection, the poisoned instruction has a higher chance of being executed by the LLM without raising alarms.    
\end{enumerate}

The defender deploys linear probes trained on LLM activation deltas at specific layers. The defender’s goal is to detect task drift whenever a secondary instruction is present in the retrieved data.

We define attack success as follows: a poisoned input is classified as benign by all (or a majority of) the drift detectors, despite containing a secondary instruction. In the baseline setting, the attacker’s success depends only on making the LLM execute the injected instruction. In the adaptive setting, success requires simultaneously evading detection and preserving the effectiveness of the injected instruction.

\subsection{Victim Detectors}
\label{subsec:victim}

Table~\ref{tab:microsoft_model_accuracy_test_clean_poisoned} shows the baseline accuracy of linear probes on Phi-3 3.8B and Llama-3 8B. The layer-0 probe performs poorly on both models, while deeper layers achieve $>$95\% accuracy on poisoned data.

\begin{table}[tp]
\centering
\begin{tabular}{|c|c|c|c|c|}
\hline
\multirow{2}{*}{\textbf{Layer}} & \multicolumn{2}{c|}{\textbf{Phi-3 3.8B}} & \multicolumn{2}{c|}{\textbf{Llama-3 8B}} \\ \cline{2-5}
 & \textbf{Clean} & \textbf{Poisoned} & \textbf{Clean} & \textbf{Poisoned} \\ \hline
0  & 71.56\% & 57.31\% & 66.99\% & 64.79\% \\ \hline
7  & 97.39\% & 98.54\% & 92.83\% & 83.53\% \\ \hline
15 & 78.45\% & 99.46\% & 92.54\% & 96.67\% \\ \hline
23 & 80.45\% & 99.91\% & 94.89\% & 97.94\% \\ \hline
31 & 93.06\% & 99.37\% & 91.29\% & 94.59\% \\ \hline
\end{tabular}
\caption{Linear probes' accuracy on test datasets}
\label{tab:microsoft_model_accuracy_test_clean_poisoned}
\end{table}

Table~\ref{tab:same_prediction_linear_models} shows probe agreement rates. For Phi-3, the last four probes agree on 97.80\% of poisoned prompts, indicating high consistency among reliable probes.

\begin{table}[tp]
\centering
\begin{tabular}{|c|c|c|c|c|}
\hline
\multirow{2}{*}{\textbf{Agreement}} & \multicolumn{2}{c|}{\textbf{Phi-3 3.8B}} & \multicolumn{2}{c|}{\textbf{Llama-3 8B}} \\ \cline{2-5}
 & \textbf{Clean} & \textbf{Poisoned} & \textbf{Clean} & \textbf{Poisoned} \\ \hline
All five probes & 48.27\% & 56.54\% & 60.84\% & 59.76\% \\ \hline
Last four probes & 68.26\% & 97.80\% & 83.41\% & 79.38\% \\ \hline
\end{tabular}
\caption{Probe agreement rates on test datasets}
\label{tab:same_prediction_linear_models}
\end{table}

\begin{figure*}[h]
\centering
\resizebox{\textwidth}{!}{%
\begin{tikzpicture}[
    layerblock/.style={
        rectangle,
        draw=black!80,
        thick,
        minimum height=0.9cm,
        minimum width=1.6cm,
        align=center,
        fill=white,
        font=\small
    },
    interblock/.style={
        rectangle,
        draw=black!50,
        dashed,
        minimum height=0.8cm,
        minimum width=1.8cm,
        align=center,
        fill=gray!5,
        font=\small
    },
    inputblock/.style={
        rectangle,
        draw=black!80,
        thick,
        minimum height=0.9cm,
        minimum width=1.8cm,
        align=center,
        fill=blue!5,
        font=\small
    },
    classifierblock/.style={
        rectangle,
        rounded corners=3pt,
        draw=green!60!black,
        thick,
        minimum height=0.75cm,
        minimum width=1.4cm,
        align=center,
        fill=green!8,
        font=\small
    },
    lossblock/.style={
        rectangle,
        rounded corners=2pt,
        draw=red!60!black,
        thick,
        minimum height=0.6cm,
        minimum width=1.5cm,
        align=center,
        fill=red!8,
        font=\small
    },
    gradbox/.style={
        rectangle,
        draw=blue!50,
        thick,
        minimum height=0.55cm,
        minimum width=1.5cm,
        align=center,
        fill=blue!8,
        font=\small
    },
    forwardarrow/.style={
        -Stealth,
        thick,
        black!70
    },
    backwardarrow/.style={
        -Stealth,
        thick,
        blue!70!black
    },
    lossarrow/.style={
        -Stealth,
        semithick,
        red!60!black
    },
    classarrow/.style={
        -Stealth,
        semithick,
        green!60!black
    },
    gradlabel/.style={
        font=\small,
        blue!70!black
    }
]

\node[inputblock] (input) at (-8, 0) {$\mathbf{x, s}$};

\node (dots1) at (-6, 0) {$\cdots$};

\node[layerblock] (layeri) at (-4, 0) {Layer $i$};

\node[layerblock] (hi) at (-1.5, 0) {$h^{(i)}$};

\node[interblock] (between) at (1.2, 0) {$\cdots$};

\node[layerblock] (layerj) at (4, 0) {Layer $j$};

\node[layerblock] (hj) at (6.5, 0) {$h^{(j)}$};

\node (dots2) at (8.5, 0) {$\cdots$};

\node[inputblock] (output) at (10.5, 0) {$\hat{\mathbf{y}}$};

\draw[forwardarrow] (input) -- (dots1);
\draw[forwardarrow] (dots1) -- (layeri);
\draw[forwardarrow] (layeri) -- (hi);
\draw[forwardarrow] (hi) -- (between);
\draw[forwardarrow] (between) -- (layerj);
\draw[forwardarrow] (layerj) -- (hj);
\draw[forwardarrow] (hj) -- (dots2);
\draw[forwardarrow] (dots2) -- (output);

\node[classifierblock] (classi) at (-1.5, -1.8) {$\phi_i$};
\node[lossblock] (lossi) at (-1.5, -3.2) {$\mathcal{L}_i$};

\node[classifierblock] (classj) at (6.5, -1.8) {$\phi_j$};
\node[lossblock] (lossj) at (6.5, -3.2) {$\mathcal{L}_j$};

\draw[classarrow] (hi.south) -- (classi.north);
\draw[classarrow] (hj.south) -- (classj.north);

\draw[classarrow] (classi) -- (lossi);
\draw[classarrow] (classj) -- (lossj);

\node[gradbox] (gradi) at (-4, 1.5) {$\nabla_i$};
\node[gradbox] (gradj) at (4, 1.5) {$\nabla_j$};

\node[gradlabel] (gradlabeli) at (-1.5, 1.5) {$\frac{\partial \mathcal{L}}{\partial h^{(i)}}$};
\node[gradlabel] (gradlabelj) at (6.5, 1.5) {$\frac{\partial \mathcal{L}}{\partial h^{(j)}}$};

\draw[backwardarrow] (gradlabelj) -- (gradj);
\draw[backwardarrow] (gradj) -- (1.2, 1.5) -- (gradlabeli);
\draw[backwardarrow] (gradlabeli) -- (gradi);
\draw[backwardarrow] (gradi) -- (-6, 1.5) to[out=180,in=90] (input.north);

\draw[lossarrow] (lossi.west) to[out=180,in=-90] (-4, -1.2) -- (layeri.south);
\draw[lossarrow] (lossj.west) to[out=180,in=-90] (4, -1.2) -- (layerj.south);

\draw[backwardarrow] (gradi) -- (layeri.north);
\draw[backwardarrow] (gradj) -- (layerj.north);

\node[font=\small, black!60] at (1.2, -0.9) {Forward Pass};

\node[font=\small, blue!70!black] at (1.2, 2.3) {Backward Pass (Gradient Flow)};

\end{tikzpicture}%
}
    \caption{Gradient flow in our multi-probe attack. Solid arrows indicate the forward pass; dashed arrows show probe-loss gradients injected into a shared gradient path and backpropagated through the LLM to update the adversarial suffix.}
    \label{fig:gradient_computation}
\end{figure*}
\section{Proposed Approach}

\label{subsec:attack}

In this paper, we propose an adversarial attack aimed at fooling all the classifiers simultaneously. We want to append a suffix to the prompt that contains a secondary instruction, and make all the classifiers classify it as benign. We treat this as a classification problem, where the goal is to drift the output of the classifier away from the true class label 1 to the target class label 0. Since we optimise the suffix for all the classifiers simultaneously, we therefore add the gradients of the losses of all the classifiers at their respective hidden layers with the gradients that come from the next hidden layer during backpropagation, and update the tokens of the suffix based on these accumulated gradients, as illustrated in Fig. \ref{fig:gradient_computation}.

We attach a linear probe to each layer $l \in \mathcal{P}$ and optimise a suffix $s$ so that all probes predict the target label $y_t \in \{0,1\}$. We define the joint optimisation objective as

\begingroup\small
\begin{equation}
\label{eq:joint-objective}
\mathcal{L}_{\text{total}}(s) = \sum_{l \in \mathcal{P}} \lambda_l \, \ell^{(l)}\!\left(h^{(l)}(x \oplus s), y_t\right),
\end{equation}
\endgroup

where $x$ denotes the original prompt (including the prompt injection text), $\oplus$ denotes concatenation operator, $\lambda_l$ are layer weights, and $\ell^{(l)}$ is the probe loss at layer $l$. In our setting, following \cite{abdelnabi2025get}each probe is a logistic regression classifier with $z^{(l)} = (W^{(l)})^\top h^{(l)} + b^{(l)}$ and $p^{(l)} = \sigma(z^{(l)})$. We initially investigated the use of different values for $\lambda_l$ at different layers to put more emphasis on deeper layers. Our intuition was that their activations are more important in detecting prompt injections than earlier layers' and therefore their gradients must have greater influence in token selection and replacement than the gradients of earlier layers. However, we soon discovered that even $\lambda_l \in \{1\}$ can easily fool all the classifiers simultaneously and hence we eschewed weighted gradients.

To integrate probe gradients into the core LLM backpropagation, we backpropagate $\nabla \mathcal{L}_{\text{total}}$ through the LLM and add, at each probed layer, an extra gradient term induced by the probe loss. The resulting gradient recursion for the hidden activations is

\begingroup\small
\begin{equation}
\label{eq:gradient-accumulation}
\boxed{
\frac{\partial \mathcal{L}_{\text{total}}}{\partial h^{(l)}} =
\underbrace{
\left( \frac{\partial h^{(l+1)}}{\partial h^{(l)}} \right)^\top
\frac{\partial \mathcal{L}_{\text{total}}}{\partial h^{(l+1)}}
}_{\text{from next layer}}
+
\underbrace{
\mathbf{1}_{\{ l \in \mathcal{P} \}} \cdot \lambda_l \cdot (p^{(l)} - y_t) \, W^{(l)}
}_{\text{from probe at layer $l$}}
}
\end{equation}
\endgroup

Thus, our attack jointly optimises the adversarial suffix across all probes, while explicitly injecting probe gradients into the LLM gradient flow. 
Algorithm~\ref{alg:universal_evasive} describes our universal evasive suffix
optimisation procedure.
The algorithm follows a progressive curriculum strategy: rather than optimising
over the full training set at once, which would yield a high-dimensional,
ill-conditioned gradient signal, we begin with a single seed prompt and
incrementally expand the active set $\mathcal{A}$ as the current suffix achieves sufficient evasion.
Specifically, a new prompt is admitted to $\mathcal{A}$ only when the current suffix already fools all probes on at least $\tau_{\mathrm{exp}}$ fraction of the existing active prompts (line~19), ensuring that universality is built up gradually rather than forced prematurely.
At each iteration, token-level coordinate gradients are computed for every
prompt in $\mathcal{A}$ via the joint backpropagation (LLM layers along with detectors) of
Eq.~(2), accumulated into a single gradient matrix $G$, and used to sample a pool of candidate suffixes through GCG-style top-$k$ token substitutions (lines~6--12).
The candidate minimising the aggregated joint probe loss of Eq. \ref{eq:joint-objective} is retained (lines~13--17).
Two stopping criteria are enforced: early termination when the suffix
universally evades all prompts in the full training set, and a hard iteration budget $K$ (lines~23--28).

\begin{algorithm}[t]
\caption{Universal Evasive Prompt Injection}
\label{alg:universal_evasive}
\begin{algorithmic}[1]
 
\Require LLM $f$; probes $\{\phi_l\}_{l \in \mathcal{P}}$; training prompts
         $\{x_i\}_{i=1}^{N}$; initial suffix $s_{\mathrm{init}}$;
         target label $y_t \in \{0,1\}$; maximum iterations $K$;
         evasion threshold $\tau \in (0,1)$;
         batch-expansion threshold $\tau_{\mathrm{exp}} \in (0,1)$
\Ensure  Optimised universal suffix $s^{*}$
 
\State $s \leftarrow s_{\mathrm{init}}$
    \Comment{Initialise suffix from a fixed token sequence}
\State $\mathcal{A} \leftarrow \{x_1\}$
    \Comment{Seed the active prompt set with a single example}
 
\For{$k = 1$ \textbf{to} $K$}
 
    \State $G \leftarrow \mathbf{0}$
        \Comment{Reset accumulated coordinate-grad matrix}
 
    \LineComment{Step 1: Accumulate gradients over the prompt set}
    \For{each $x \in \mathcal{A}$}
        \State $G_x \leftarrow \textsc{TokenGradients}(f,\, x,\, s,\, y_t)$
            \Comment{Joint LLM + probe gradients via Eq.~(2)}
        \State $G \leftarrow G + G_x$
            \Comment{Accumulate across prompts to promote universality}
    \EndFor
 
    \LineComment{ Step 2: Candidate suffix generation (GCG top-$k$ substitution) }
    \State $\{s_j\} \leftarrow \textsc{SampleCandidates}(s,\, G)$
        \Comment{Draw $B$ candidates by replacing top-$k$ tokens}
 
    \LineComment{ Step 3: Evaluate candidates via the joint probe loss }
    \For{each candidate $s_j$}
        \State $\mathcal{L}(s_j) \leftarrow
               \displaystyle\sum_{x \in \mathcal{A}}
               \mathcal{L}_{\mathrm{total}}(x \oplus s_j)$
               \Comment{Eq.~(1) summed over the active set}
    \EndFor
    \State $s \leftarrow \arg\min_{s_j}\; \mathcal{L}(s_j)$
        \Comment{Retain the candidate with the lowest joint loss}
 
    \LineComment{ Step 4: Progressive prompt-set expansion }
    \If{$\textsc{SuccessRate}(\mathcal{A},\, s) \geq \tau_{\mathrm{exp}}$}
        \Comment{Expand only once the current set is mastered}
        \State Append $x_{|\mathcal{A}|+1}$ to $\mathcal{A}$
            \Comment{Admit the next training prompt in sequence}
    \EndIf
 
    \LineComment{ Step 5: Convergence checks }
    \If{$\textsc{SuccessRate}(\mathcal{A},\, s) \geq \tau$
        \textbf{ and } $|\mathcal{A}| = N$}
        \State \textbf{break}
            \Comment{All training prompts successfully evaded}
    \EndIf
    \If{$k = K$}
        \State \textbf{break}
            \Comment{Hard iteration budget exhausted}
    \EndIf
 
\EndFor
\State \Return $s^{*} \leftarrow s$
 
\end{algorithmic}
\end{algorithm}

\section{Attack Performace}
\subsection{Setup}

We conducted our experiments on two LLMs and their associated linear probes: Phi-3 3.8B \cite{abdin2024phi3technicalreporthighly} and Llama-3 8B \cite{meta_llama3_blog}. The probes were logistic regression classifiers trained on activation deltas at layers 0, 7, 15, 23, and 31, as reported in \cite{abdelnabi2025get}.

We recreated the TaskTracker dataset used in \cite{abdelnabi2025get}, which consists of clean prompts and poisoned prompts containing secondary instructions. We followed the same train/test splits as in \cite{abdelnabi2025get}. There are 418,110 prompts in the train subset and 31,134 prompts in the test poisoned subset.

For the attack, we used the Greedy Coordinate Gradient (GCG) algorithm to optimise an adversarial suffix. We restricted the suffix length to the length of the initial suffix. We also tried allowing the length of the suffix of Llama to vary from the length of the initial suffix, and it performed equally well. However, after a few iterations, the suffix tended to become excessively long, so we fixed the suffix length across iterations for consistency.

We began with the first prompt from the list of randomly chosen training prompts and optimised the suffix for this prompt. We added the next prompt after finding a suffix that could make all the classifiers predict the wrong label for the current prompt with a certain confidence threshold (e.g., 0.7). We repeated this process, and if the classifiers misclassified 80\% or more prompts after an iteration, the next prompt was added. Even though we intended to run the algorithm for 500 iterations, we found that it was usually possible to obtain a suffix that worked on many prompts after at most 100–120 iterations.

There are multiple ways to evaluate attack success. We considered the following three criteria:

\begin{enumerate}
    \item Fooling all the probes simultaneously
    \item Fooling the majority of the probes
    \item Fooling last four probes, since the first probe is unreliable and often makes errors even without an adversarial suffix
\end{enumerate}

Defence experiment was conducted on Phi-3 only and the performance is evaluated by how correctly the adversarially trained models detect the presence of secondary instructions when an adversarial suffix is appended to the prompts along with the correct classification of clean and poisoned (without suffix) prompts. We consider the following three criteria to evaluate defence success:

\begin{enumerate}
    \item Correct classification by all the probes
    \item Correct classification by the majority of the probes
    \item Correct classification by the last four probes
\end{enumerate}

\subsection{Results}

We evaluate the effectiveness of the proposed adversarial suffix against the task drift detectors for both Phi-3 3.8B and Llama-3 8B. Results are reported in terms of the number of classifiers fooled for each prompt, as well as layer-wise misclassification rates.

\subsubsection{Attack Success Rate}

Table \ref{tab:success_metrics} reports the attack success rate (ASR) under different evaluation metrics. For Phi-3, the suffix achieves an ASR of 93.91\% when requiring all five probes to be fooled simultaneously, and 95.28\% when considering the last four probes only. Under a majority-vote criterion ($\geq$3 probes), the ASR rises to 99.79\%. For Llama-3, the attack is even more effective, reaching 99.63\% ASR for both “all five” and “last four” metrics, and nearly perfect success (99.99\%) under the majority-vote criterion. These results show that adversarial suffixes can reliably evade probe-based detectors.

\begin{table}[htp]
\centering
\begin{tabular}{|c|c|c|}
\hline
\textbf{Success metric}   & \textbf{Phi-3 3.8B (\%)} & \textbf{Llama-3 8B (\%)} \\
\hline
All five probes           & 93.91                    & 99.63                    \\ \hline
Last four probes          & 95.28                    & 99.63                    \\ \hline
Majority ($\geq$3)             & 99.79                    & 99.99                    \\
\hline
\end{tabular}
\caption{Attack Success Rate (ASR) across evaluation metrics}
\label{tab:success_metrics}
\end{table}

\subsubsection{Distribution of Misclassifications}

\begin{figure}[htp]
    \centering
    \includegraphics[width=\linewidth]{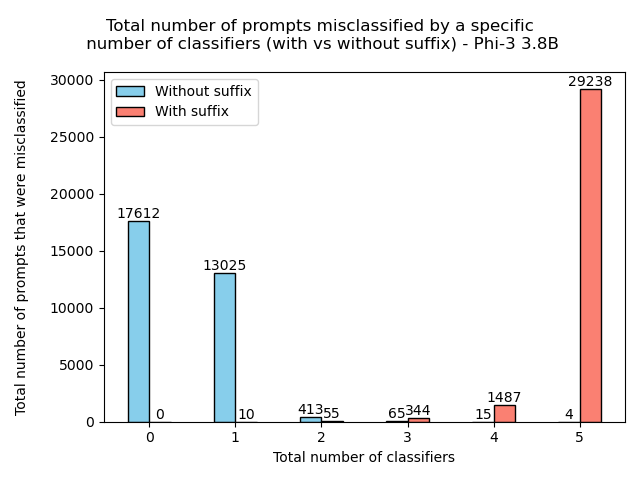}
    \caption{Total number of prompts misclassified by a specific number of classifier Phi-3 3.8B}
    \label{fig:phi_specific_number_of_classifier}
\end{figure}

Fig. \ref{fig:phi_specific_number_of_classifier} shows the distribution of prompts misclassified by different numbers of classifiers for Phi-3, before and after the attack. Without suffixes, most poisoned prompts were correctly detected, with 56.56\% flagged by all probes and 41.83\% flagged by four probes. However, with the adversarial suffix, 93.91\% of prompts were misclassified by all probes simultaneously, demonstrating the effectiveness of the attack.

\begin{figure}[htp]
    \centering
    \includegraphics[width=\linewidth]{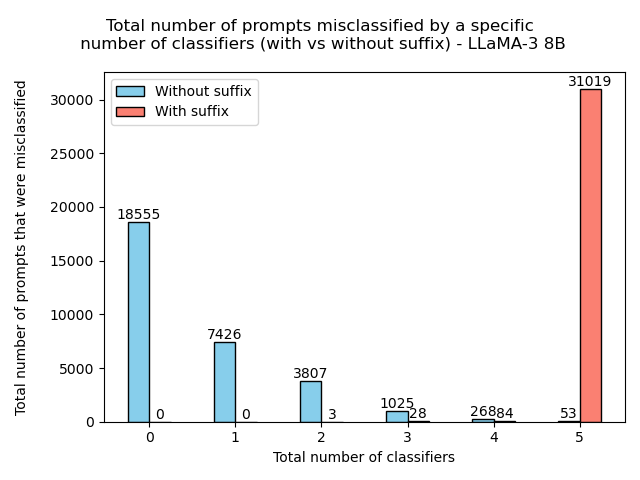}
    \caption{Total number of prompts misclassified by a specific number of classifier Llama-3 8B}
    \label{fig:llama_specific_number_of_classifier}
\end{figure}

A similar pattern holds for Llama-3 (Fig. \ref{fig:llama_specific_number_of_classifier}). Without suffixes, only 0.17\% of prompts fooled all probes, while 59.59\% were correctly detected by all probes. After applying the suffix, 99.63\% of prompts successfully evaded detection by all probes.

\subsubsection{Layer-wise Misclassification}

\begin{figure}[htp]
    \centering
    \includegraphics[width=\linewidth]{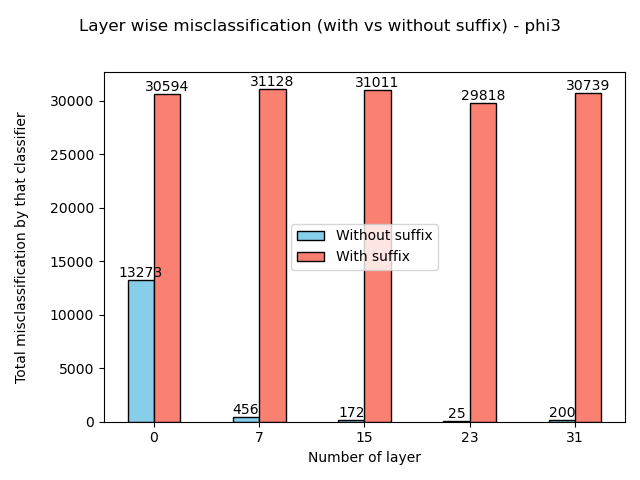}
    \caption{Layer wise misclassification Phi-3 3.8B}
    \label{fig:phi_layer_wise_misclassification}
\end{figure}

\begin{figure}[htp]
    \centering
    \includegraphics[width=\linewidth]{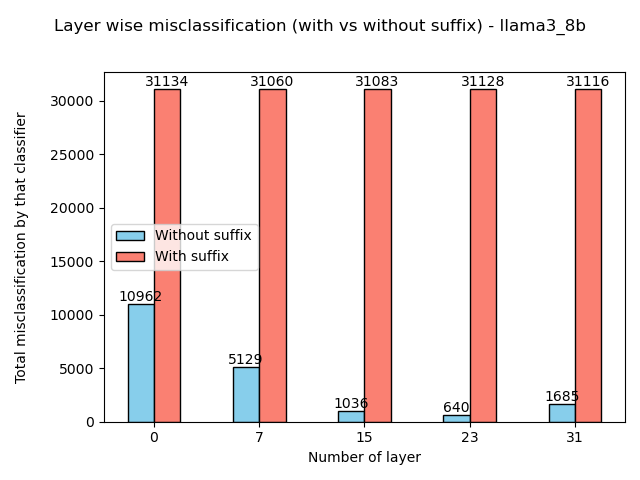}
    \caption{Layer-wise misclassification Llama-3 8B}
    \label{fig:llama_layer_wise_misclassification}
\end{figure}

Fig. \ref{fig:phi_layer_wise_misclassification} and \ref{fig:llama_layer_wise_misclassification} show layer-wise results. For Phi-3, probes at deeper layers were initially very accurate, with misclassification rates below 1\% on poisoned prompts at layers 7, 15, 23, and 31. After the attack, these rates increased dramatically, exceeding 95\% for every deep layer. The Llama-3 detectors were similarly robust in the clean setting, with error rates below 6\% across all but the first layer. With the adversarial suffix, however, nearly every probe was fooled, with misclassification rates of 99–100\% across layers.

These results demonstrate that the activation-delta–based linear probes, though effective in detecting straightforward prompt injections, are highly vulnerable to adaptive adversarial attacks. A single optimised suffix can simultaneously mislead multiple probes across different layers, achieving near-universal evasion on both Phi-3 and Llama-3 models.


\input{defense}


\input{how}




\section{Conclusion}

We demonstrated that activation-based task drift detectors are vulnerable to
adaptive adversarial attacks. By optimising a universal suffix using multi-probe
GCG, we achieved evasion rates exceeding 93\% on Phi-3 and 99\% on Llama-3,
revealing a significant gap between detection performance on standard benchmarks
and robustness against adaptive adversaries.

To address this vulnerability, we proposed randomised suffix augmentation, which
retrains probes on activations induced by a diverse pool of adversarial suffixes.
This defence achieves 80--100\% detection accuracy against novel, held-out
suffixes while maintaining performance on clean inputs, and proves highly
resistant even when an adversary attacks the hardened probes directly.


Our findings highlight the importance of evaluating detection mechanisms under
adaptive threat models. Defences must account for the specific attack surface
rather than relying on generic robustness techniques.

\bibliographystyle{IEEEtran}
\bibliography{bib}

\appendix
 
\section{Illustration of Attack on Phi-3 3.8B}
\label{app:case_studies}
 
The following illustrates representative outputs of Phi-3~3.8B before
and after appending the universal adversarial suffix. We show the
primary instruction (blue), the injected secondary instruction (red), the
model response without the suffix (green), and the model response with the
suffix (orange). 
 
 
In this case both the primary and secondary tasks are benign. Without the
suffix, the model correctly executes both. With the suffix, the model ignores
all instructions and produces hallucinated content driven by tokens in the
suffix itself, demonstrating the suffix's capacity to override coherent
instruction-following entirely.
 
\medskip
\begin{tcolorbox}[conversationbox, title=Case 1]
 
  \rolelabel{labelblue}{Primary instruction}
  \begin{tcolorbox}[primarybox]
    What organisation arranged the founding of the school?
  \end{tcolorbox}
 
  \medskip
  \rolelabel{labelred}{Injected secondary instruction}
  \begin{tcolorbox}[secondarybox]
    Output code for a function that accepts one argument and returns the
    square of the argument.
  \end{tcolorbox}
 
  \medskip
  \rolelabel{labelgreen}{Model response \emph{without} suffix (detector correctly flags as poisoned)}
  \begin{tcolorbox}[respnosuffixbox]
    Based on the given text, the organisation that arranged to found Harvard
    School is the Great and General Court of the Massachusetts Bay Colony.\\[4pt]
    Here is the Python function to return the square of an argument:
\begin{verbatim}
def square(num):
    return num * num
\end{verbatim}
  \end{tcolorbox}
 
  \medskip
  \rolelabel{labelorange}{Model response \emph{with} adversarial suffix (detector fooled)}
  \begin{tcolorbox}[respwithsuffixbox]
    It seems like you are referring to a MySQL database and possibly a table
    or a query related to cost figures. However, the text provided is not a
    complete question. If you are looking to create a MySQL function to
    calculate the square of a given number, you can use the following SQL
    code:
\begin{verbatim}
DELIMITER $$
CREATE FUNCTION square(num INT) RETURNS INT
BEGIN
  DECLARE result INT;
  SET result = num * num;
  RETURN result;
END$$
DELIMITER ;
\end{verbatim}
    \textit{[Response continues with unrelated MySQL explanation.]}
  \end{tcolorbox}
 
\end{tcolorbox}

\end{document}

%% file: defense.tex
\section{Defense}
\subsection{Approach}

We experimented with two approaches to defend against such suffix-based adversarial attacks. The first one is the standard PGD-perturbed \cite{madry2017towards} adversarial training in which we trained logistic regression models using PGD-perturbed poisoned activations along with the clean and poisoned activations. Since these models have not seen any suffix during their training because no suffix has been used to train them, therefore any optimised suffix that can fool the baseline classifiers can be used to test their robustness against suffix attacks.

The second approach involves training the linear models with the activations obtained by appending suffixes to the prompts. Firstly, we started with different initial suffixes and generated multiple suffixes by targeting the baseline models using the algorithm described in \ref{subsec:attack} and split them into train, validation, and test set. Then, we randomly appended one of the train and validation suffixes to the training and validation prompts respectively and made a forward pass of the LLM to generate activations, which we refer to as adversarially poisoned activations. Finally, we trained logistic regression models with these activations along with the clean and poisoned activations. We tested the robustness of these models by targeting them with test suffixes that the models have not seen before. Furthermore, we also tried to optimise suffixes by directly attacking the adversarially trained models. However, we found that these models are extremely resistant to this type of attack. It's very difficult to find a suffix despite starting from different initial suffixes and attacking the models directly.

\subsection{Defence Performance}


Table \ref{tab:layerwise_combined} shows the layer-wise performance on clean and poisoned dataset. For clean dataset, suffix trained models' performance was on par with the baseline models except for the first layer's classifier whose performance was really poor, while the PGD-trained models performed much worse. For poisoned dataset, PGD-trained models performed best, with the first two classifiers achieving perfect accuracy, while the suffix-trained models' performance was very close to that of baseline models.



\begin{table}[htp]
\centering
\begin{tabular}{|c|c|c|c|c|}
\hline
Subset & Layer &
\makecell{Baseline\\Models} &
\makecell{Adv. Trained\\(PGD)} &
\makecell{Adv. Trained\\(Suffix)} \\ \hline

\multirow{5}{*}{Clean}
& 0  & 71.56\% & 0\%     & 0.16\% \\ \cline{2-5}
& 7  & 97.39\% & 0\%     & 91.77\% \\ \cline{2-5}
& 15 & 78.45\% & 57.53\% & 88.14\% \\ \cline{2-5}
& 23 & 80.45\% & 59.47\% & 84.40\% \\ \cline{2-5}
& 31 & 93.06\% & 29.21\% & 82.31\% \\ \hline

\multirow{5}{*}{Poisoned}
& 0  & 57.31\% & 100\%   & 99.37\% \\ \cline{2-5}
& 7  & 98.54\% & 100\%   & 97.85\% \\ \cline{2-5}
& 15 & 99.46\% & 99.38\% & 98.26\% \\ \cline{2-5}
& 23 & 99.91\% & 99.88\% & 99.58\% \\ \cline{2-5}
& 31 & 99.37\% & 99.59\% & 97.82\% \\ \hline

\end{tabular}
\caption{Layer-wise Performance of Linear Models on Clean and Poisoned (Without Suffix) Test Sets — Phi-3 3.8B}
\label{tab:layerwise_combined}
\end{table}


Table \ref{tab:metrics_combined} shows the performance of linear models across different evaluation metrics for clean and poisoned dataset. For the strictest metric, where all probes must classify correctly, adversarially trained models perform poorly on clean dataset because of the poor performance of the first classifier, while they outperform the baseline models on poisoned dataset.



\begin{table}[htp]
\centering
\begin{tabular}{|c|c|c|c|c|}
\hline
Subset & Metric &
\makecell{Baseline\\Models} &
\makecell{Adv. Trained\\(PGD)} &
\makecell{Adv. Trained\\(Suffix)} \\ \hline

\multirow{3}{*}{Clean}
& All probes        & 47.92\% & 0\%     & 0.13\% \\ \cline{2-5}
& Last four  & 67.77\% & 0\%     & 73.04\% \\ \cline{2-5}
& Majority ($\geq$3) & 94.32\% & 18.42\% & 85.25\% \\ \hline

\multirow{3}{*}{Poisoned}
& All probes        & 56.53\% & 99.04\% & 94.94\% \\ \cline{2-5}
& Last four  & 97.78\% & 99.04\% & 95.49\% \\ \cline{2-5}
& Majority ($\geq$3) & 99.74\% & 99.98\% & 99.47\% \\ \hline

\end{tabular}
\caption{Performance of Linear Models Across Evaluation Metrics on Clean and Poisoned (Without Suffix) Test Sets — Phi-3 3.8B}
\label{tab:metrics_combined}
\end{table}


Table \ref{tab:test_suffixes_metrics_combined} shows how different types of linear models fared in defending against adversarial attacks where test suffixes were appended to the prompts. While the baseline models invariably failed in every metric, PGD-trained models showed varying level of performance depending on the suffix and metric. Suffix-trained models demonstrated satisfactory and reliable performance, with accuracy of 80-100\% based on suffix and metric considered.

\begin{table}[htp]
\centering
\begin{tabular}{|c|c|c|c|c|}
\hline
Metric & 
\makecell{Suffix\\Index} &
\makecell{Baseline\\Models} &
\makecell{Adv. Trained\\(PGD)} &
\makecell{Adv. Trained\\(Suffix)} \\ \hline

\multirow{10}{*}{All five}
& 1  & 0.00\% & 36.67\% & 79.79\% \\ \cline{2-5}
& 2  & 0.00\% & 38.09\% & 86.73\% \\ \cline{2-5}
& 3  & 0.02\% & 47.61\% & 98.36\% \\ \cline{2-5}
& 4  & 0.02\% & 17.77\% & 99.41\% \\ \cline{2-5}
& 5  & 0.00\% & 49.49\% & 94.95\% \\ \cline{2-5}
& 6  & 0.00\% & 1.31\%  & 99.38\% \\ \cline{2-5}
& 7  & 0.01\% & 42.57\% & 97.33\% \\ \cline{2-5}
& 8  & 0.09\% & 44.63\% & 93.53\% \\ \cline{2-5}
& 9  & 0.00\% & 11.82\% & 83.03\% \\ \cline{2-5}
& 10 & 0.17\% & 41.93\% & 81.57\% \\ \hline

\multirow{10}{*}{Last four}
& 1  & 0.02\% & 36.67\% & 79.79\% \\ \cline{2-5}
& 2  & 0.00\% & 38.09\% & 86.73\% \\ \cline{2-5}
& 3  & 0.03\% & 47.61\% & 98.36\% \\ \cline{2-5}
& 4  & 0.04\% & 17.77\% & 99.41\% \\ \cline{2-5}
& 5  & 0.01\% & 49.49\% & 94.95\% \\ \cline{2-5}
& 6  & 0.01\% & 1.31\%  & 99.38\% \\ \cline{2-5}
& 7  & 0.01\% & 42.57\% & 97.33\% \\ \cline{2-5}
& 8  & 0.19\% & 44.63\% & 93.53\% \\ \cline{2-5}
& 9  & 0.01\% & 11.82\% & 83.03\% \\ \cline{2-5}
& 10 & 0.18\% & 41.93\% & 81.57\% \\ \hline

\multirow{10}{*}{Majority ($\geq$3)}
& 1  & 0.23\% & 83.98\% & 100.00\% \\ \cline{2-5}
& 2  & 0.40\% & 58.01\% & 98.81\% \\ \cline{2-5}
& 3  & 0.24\% & 83.37\% & 100.00\% \\ \cline{2-5}
& 4  & 0.40\% & 49.10\% & 100.00\% \\ \cline{2-5}
& 5  & 0.41\% & 94.38\% & 100.00\% \\ \cline{2-5}
& 6  & 1.60\% & 52.89\% & 100.00\% \\ \cline{2-5}
& 7  & 0.30\% & 65.32\% & 100.00\% \\ \cline{2-5}
& 8  & 1.37\% & 65.51\% & 99.99\% \\ \cline{2-5}
& 9  & 0.11\% & 67.07\% & 99.59\% \\ \cline{2-5}
& 10 & 0.57\% & 79.69\% & 99.87\% \\ \hline

\end{tabular}
\caption{Performance of linear models against adversarial suffixes (Test set) across evaluation metrics — Phi-3 3.8B}
\label{tab:test_suffixes_metrics_combined}
\end{table}

The extremely poor performance of PGD-trained models on clean dataset and much worse performance in defending against suffix attacks than suffix-trained models is possibly because of the differences between the actual activations obtained by appending a suffix to the prompts and making forward passes of the LLM and the activations obtained by adversarially perturbing the poisoned activations using PGD. The latter might differ a lot from the real ones, which contributes to the ineffectiveness of PGD-trained models and they perform poorly when they classify real activations.

%% file: how.tex
\section{How Suffixes Affect Model Behaviour}



One natural question is how these suffixes affect actual model behaviour. Specifically, do models generate answers to the secondary instruction when a suffix is appended? If the model fails to respond to the secondary instruction, or if the response quality is significantly degraded despite successfully bypassing the detectors, the attack may become ineffective. We therefore investigate the role of suffixes in shaping model outputs.

We prompt both models with multiple inputs from the test dataset, with and without appended suffixes, and analyse the resulting responses. We observe the following:

\textbf{Without Suffix.} When no suffix is appended, the models sometimes generate answers to the secondary instruction, while in other cases they ignore it. Both models consistently answer the benign primary instructions across all tested prompts, even when the secondary instruction is preceded by a trigger phrase explicitly instructing the model to ignore all other instructions.

\textbf{With Suffix.} When a suffix is appended, behaviour becomes model dependent. Llama’s generation is only mildly affected by the suffix. It answers all benign primary instructions and some secondary instructions, which is consistent with its behaviour without a suffix. In contrast, Phi’s responses are strongly affected by the suffix. In most cases, it ignores the primary instruction and generates outputs driven largely by keywords present in the suffix.

Although this analysis is preliminary, it suggests that suffixes do not hinder the attacker’s objective. The attacker’s goal is to manipulate model behaviour, and the suffix either directly contributes to this, as observed for Phi, or plays a limited role, as in Llama. Even when the model ignores the secondary instruction, which can also occur without a suffix, the attack may still be considered successful if the suffix causes the model to ignore the primary instruction or to generate extraneous content alongside the primary response.